\documentstyle[12pt]{article}
\newfont{\ffont}{msym10}                        %%
\newcommand{\beq}{\begin{equation}}             %%
\newcommand{\eeq}{\end{equation}}               %%
\newcommand{\bqry}{\begin{eqnarray}}            %%
\newcommand{\eqry}{\end{eqnarray}}              %%
\newcommand{\bqryn}{\begin{eqnarray*}}          %%
\newcommand{\eqryn}{\end{eqnarray*}}            %%
\newcommand{\preprint}[1]{\begin{table}[t]      %%
            \begin{flushright}                  %%
            \begin{large}{#1}\end{large}        %%
            \end{flushright}                    %%
            \end{table}}                        %%
\newcommand{\PD}[2]                             %%
    {\frac{\partial^{#2}}{\partial #1^{#2}}}    %%
\catcode`\@=11 \@addtoreset{equation}{section}  %%
\renewcommand{\theequation}                     %%
         {\arabic{section}.\arabic{equation}}   %%
\begin{document}
\preprint{LA-UR-96-1337 \\ IASSNS-96/34 }
\title{Mass - Proper Time Uncertainty Relation \\ in a Manifestly
Covariant
 \\ Relativistic Statistical Mechanics}
\author{\\ L. Burakovsky\thanks{Bitnet: BURAKOV@QCD.LANL.GOV} \
\\  \\  Theoretical Division, T-8 \\  Los Alamos National Laboratory \\ Los
Alamos NM 87545, USA \\  \\
L.P. Horwitz\thanks{Bitnet: HORWITZ@SNS.IAS.EDU. On sabbatical leave from
School of Physics and Astronomy, Tel Aviv University, Ramat Aviv, Israel.
Also at Department of Physics, Bar-Ilan University, Ramat-Gan, Israel  } \
\\  \\
School of Natural Sciences \\ Institute for Advanced Study \\ Princeton NJ
 08540, USA \\  \\ and \\  \\ W.C. Schieve\thanks{Bitnet: WCS@MAIL.UTEXAS.EDU}\
\\  \\ Ilya Prigogine Center \\ for Studies in Statistical Mechanics \\
University of Texas at Austin \\ Austin TX 78712, USA \\}
\date{18 April, 1996}
\maketitle
\begin{abstract}
We prove the uncertainty relation $T_{\triangle V}\triangle
m\stackrel{>}{\sim }2\pi \hbar /c^2,$ which is realized on a
statistical mechanical level for an ensemble of events in
$(1+D)$-dimensional spacetime with motion parametrized by an invariant
``proper time'' $\tau ,$ where $T_{\triangle V}$ is the average
passage interval in $\tau $ for the events which pass through a small
(typical) $(1+D)$-volume $\triangle V,$ and $\triangle m$ is the
dispersion of mass around its on-shell value in such an ensemble. We
show that a linear mass spectrum is a completely general property of
a $(1+D)$-dimensional off-shell theory.
\end{abstract}

\bigskip
\section{Introduction}
In this paper we shall use a manifestly covariant form of statistical
mechanics which has more general structure than the standard forms
of relativistic statistical mechanics, but which reduces to those theories
in a certain limit, to be described precisely below. These theories,
which are characterized classically by mass-shell constraints, and the
use, in quantum field theory, of fields which are constructed on the
basis of on-mass-shell free fields, are associated with the
statistical treatment of {\it world lines} and hence, considerable
coherence (in terms of the macroscopic structure of whole world lines
as the elementary objects of the theory) is implied. In
nonrelativistic statistical mechanics, the elementary objects of the
theory are points.
 The relativistic  analog of this essentially structureless foundation for a statistical theory is the set of points in spacetime, i.e., the so-called {\it events}, not the world lines (Currie, Jordan and Sudarshan \cite {Sudarshan} have discussed the difficulty of constructing a relativistic mechanics on the basis of world lines).

The mass of particles in a mechanical theory of events is necessarily a dynamical variable, since the classical phase space of the relativistic set of events consists of the spacetime and energy-momentum coordinates $\{{\bf q}_i, t_i;\  {\bf p}_i,E_i\}$, with no {\it a priori} constraint on the relation between the ${\bf p}_i$ and the $E_i$, and hence  such theories are ``off-shell''. It is well known from the work of Newton and Wigner \cite{NW} that on-shell relativistic quantum theories such as those governed by Klein-Gordon or Dirac type equations do not provide local descriptions (the wave functions corresponding to localized particles are  spread out); for such theories the notion of ensembles over local initial conditions is difficult to formulate. The off-shell theory that
we shall use here is,  however, precisely local in both its first and
second quantized forms \cite{AH,Shnerb}.

We finally remark that the standard formulations of  quantum relativistic
statistical mechanics, and quantum field theory at finite temperature, lack
manifest covariance on a fundamental level. As for nonrelativistic statistical
mechanics, the partition function is described by the Hamiltonian, which is not
an invariant object, and hence thermodynamic mean values do not have tensor
properties. [One could consider the invariant $p_\mu n^\mu$ in place of the
Hamiltonian \cite{Synge}, where $n^\mu$ is a unit four-vector; this construction (supplemented by a spacelike vector othogonal to $n^\mu$) implies an induced
representation for spacetime. The quantity that takes the place of the
parameter $t$ is then $x_\mu n^\mu$; in the corresponding quantum mechanics,
the space parts of (induced form of) the momentum do not commute with
this time variable. Some of the problems associated with this construction are closely related to those pointed out by Currie, Jordan and Sudarshan \cite{Sudarshan}, for which
different world lines are predicted dynamically by the change in the form of the effective Hamiltonian in different frames.] Since the form of such a theory is not constrained by covariance requirements, its dynamical structure
and predictions may be different than for a theory which satisifies these
requirements. For example, the canonical distribution of Pauli \cite{Pauli}
for the free Boltzmann gas has a high temperature limit in which the energy is
given by $3k_BT$, which does not correspond to  any known
equipartition rule, but for the corresponding distribution for the
manifestly covariant theory, the limit is $2k_BT$, corresponding to
${1\over 2}k_BT$ for each of the four relativistic degrees of freedom
\cite{HSP,BH1}. For the quantum field theories at finite temperature,
the path integral formulation \cite {Kap1} replaces the Hamiltonian in
the canonical exponent by the Lagrangian due to the infinite product of factors $\langle\phi \vert \pi\rangle$ (transition matrix element of the
canonical field and its conjugate required to give a Weyl ordered Hamiltonian
its numerical value). However, it is the $t$ variable which is analytically
continued to construct the finite temperature canonical ensemble, completely
removing the covariance of the theoretical framework. One may argue that some
frame has to be chosen for the statistical theory to be developed, and perhaps
even for temperature to have a meaning, but as we have remarked above, the
requirement of relativistic covariance has dynamical consequences (note that the model Lagrangians used in the non-covariant formulations are established with
the criterion of relativistic covariance in mind), and we argue that the choice
of a frame, if necessary for some physical reason, such as the definition  and
measurement of temperature, should be made in the framework of a manifestly
covariant  structure.

In the framework of a manifestly covariant relativistic statistical
mechanics, the dynamical evolution of a system of $N$ particles, for the
classical case, is governed by equations of motion that are of the form
of Hamilton equations for the motion of $N$ $events$ which generate the
space-time trajectories (particle world lines) as functions of a
continuous Poincar\'{e}-invariant parameter $\tau $ \cite{Stu,HP}, usually 
referred to as a ``proper time''. These events are characterized by
their positions $q^\mu =(t,{\bf q})$ and energy-momenta $p^\mu =(
E,{\bf p})$ in an $8N$-dimensional phase-space. For the quantum case, the
system is characterized by the wave function $\psi _\tau (q_1,q_2,\ldots
,q_N)\in L ^2(R^{4N}),$ with the measure $d^4q_1d^4q_2\cdots d^4q_N\equiv
d^{4N}q,$ $(q_i\equiv q_i^\mu ;\;\;\mu =0,1,2,3;\;\;i=1,2,\ldots ,N),$
describing the distribution of events, which evolves with a generalized
Schr\"{o}dinger equation \cite{HP}. The collection of events (called
``concatenation'' \cite{AHL}) along each world line corresponds to a
{\it particle,} and hence, the evolution of the state of the $N$-event
system describes, {\it a posteriori,} the history in space and time of
an $N$-particle system.

For a system of $N$ interacting events (and hence, particles) one takes
\cite{HP}
\beq
K=\sum _i\frac{p_i^\mu p_{i\mu }}{2M}+V(q_1,q_2,\ldots ,q_N),
\eeq
where $M$ is a given fixed parameter (an intrinsic property of the
particles), with the dimension of mass, taken to be the same for all the
particles of the system. The Hamilton equations are
$$\frac{dq_i^\mu }{d\tau }=\frac{\partial K}{\partial p_{i\mu }}=\frac{p_
i^\mu }{M},$$
\beq
\frac{dp_i^\mu }{d\tau }=-\frac{\partial K}{\partial q_{i\mu }}=-\frac{
\partial V}{\partial q_{i\mu }}.
\eeq
In the quantum theory, the generalized Schr\"{o}dinger equation
\beq
i\frac{\partial }{\partial \tau }\psi _\tau (q_1,q_2,\ldots ,q_N)=K
\psi _\tau (q_1,q_2,\ldots ,q_N)
\eeq
describes the evolution of the $N$-body wave function
$\psi _\tau (q_1,q_2,\ldots ,q_N).$ To illustrate the meaning of this
wave function, consider the case of a single free event. In this case
(1.3) has the formal solution
\beq
\psi _\tau (q)=(e^{-iK_0\tau }\psi _0)(q)
\eeq
for the evolution of the free wave packet. Let us represent $\psi _\tau
(q)$ by its Fourier transform, in energy-momentum space:
\beq
\psi _\tau (q)=\frac{1}{(2\pi )^2}\int d^4pe^{-i\frac{p^2}{2M}\tau }
e^{ip\cdot q}\psi _0(p),
\eeq
where $p^2\equiv p^\mu p_\mu ,$ $p\cdot q\equiv p^\mu q_\mu ,$ and $\psi
_0(p)$ corresponds to the initial state. Applying the Ehrenfest arguments
of stationary phase to obtain the principal contribution to $\psi _\tau
(q)$ for a wave packet at $p_c^\mu ,$ one finds ($p_c^\mu $ is the peak
value in the distribution $\psi _0(p))$
\beq
q_c^\mu \simeq \frac{p_c^\mu }{M}\tau ,
\eeq
consistent with the classical equations (1.2). Therefore,
the central peak of the wave packet moves along the classical
trajectory of an event, i.e., the classical world line.

It is clear from the form of (1.3) that one can construct  relativistic
transport theory in a form analogous to that of the nonrelativistic theory; a
relativistic Boltzmann equation and its consequences, for example, was studied
in ref. \cite{HSS}.

Since, in the Hilbert space $L^2(R^4)$ the operators $x^\mu ,p^\nu $ obey the canonical commutation relations $(g^{\mu \nu }={\rm diag}(-,+,+,+))$
\beq
[x^\mu ,p^\nu ]=i\hbar g^{\mu \nu },
\eeq
the uncertainty relations
\beq
\triangle x\;\triangle p\geq\frac{\hbar }{2},\;\;\;
\triangle t\;\triangle E\geq\frac{\hbar }{2}
\eeq
follow directly from the mathematical structure of the theory, and are
on the same footing (with the usual statistical interpretation
\cite{FW}). The dispertion $\triangle t$ is a property of the wave
function $\psi _\tau (q)$ at a given $\tau .$

Arshansky and Horwitz \cite{AH} have studied thought experiments analogous to those discussed by Landau and Peierls \cite{LP}, within the framework of the manifestly covariant relativistic quantum theory which we are using here, and derived the (causal) Landau-Peierls relation
\beq
\triangle t\;\triangle p\stackrel{>}{\sim }\frac{\hbar }{c},
\eeq
concerning the time interval $\triangle t$ during which the momentum of a particle is measured, and the momentum dispersion of the state. 

In this paper we shall discuss another uncertainty relation following
directly from the mathematical structure of the theory which is
realized on a statistical mechanical level (for an ensemble of events),
\beq
T_{\triangle V}\;\triangle m\stackrel{>}{\sim }\frac{2\pi \hbar}{c^2},
\eeq
where $\triangle m$ is the mass dispersion around the on-shell value
due to the off-shellness of the events making up the ensemble, and
$T_{\triangle V}$ is the average passage interval in $\tau $ for the
events which pass through the small (typical) four-volume $\triangle
V$ in the neighborhood of the $R^4$-point.\footnote{This result is
analogous to the {\it nonrelativistic} $\triangle t\;\triangle E$
relation, which, as for (1.10), does not follow from commutation
relations and the Schwartz inequality.  The nonrelativistic time-energy 
uncertainty relation, in fact, follows from (1.10) and (1.11) in the
  nonrelativistic 
limit for which $\tau \rightarrow t,$ $\langle E\rangle /M \rightarrow 1,$
and $\triangle m\;c^2 \rightarrow \triangle E$. This result implies
the existence, in the non-relativistic limit, of a residual ensemble
over $t$, consistently with the treatment of the non-relativistic
limit given in ref. [7].} The four-volume $\triangle V$ is the smallest that can be considered a macrovolume in representing the ensemble. $T_{\triangle V}$ is related to the (average) extent of the ensemble along the time axis, through the Hamilton equation (1.2) for $\mu =0$ (in the sense of a statistical average), 
\beq
\frac{\triangle t}{T_{\triangle V}}=\frac{\langle E\rangle }{M},
\eeq
if the ensemble is constructed with the minimum time span to
characterize the physical system.

\section{Ideal relativistic gas of events}
To describe an ideal gas of events in the grand canonical ensemble, we
use the expression for the number of events given in \cite{HSP} (in the following we shall use the system of units in which $\hbar =c=k_B=1,$ unless otherwise specified),
\beq
N=\sum _{p^\mu }n_{p^\mu }=\sum _{p^\mu }\frac{1}{e^{(E-\mu -\mu _K\frac{m^2}{2M})/T}\mp 1},
\eeq
where $E\equiv p^0,$ $m^2 \equiv  -p^2 = - p^\mu p_\mu ,$ and the sign in the denominator of (2.1) is determined by the event statistics in the usual way;
$\mu _K$ is an additional mass potential \cite{HSP}, which arises in the
grand canonical ensemble as the derivative of the free energy with respect to
the value of the dynamical evolution function $K$, interpreted as the invariant
mass of the system. In the kinetic theory \cite{HSP}, $\mu _K$ enters as a Lagrange multiplier for the equilibrium distribution for $K,$ as $\mu $ is for $N$ and $1/T$ for $E.$  In order to simplify subsequent considerations, we
shall take it to be a fixed parameter. 

We restrict ourselves, in the following, to the case of the events obeying Bose-Einstein statistics and use, therefore, the relation (2.1) with the minus sign in the denominator. 
To ensure a positive-definite value for $n_{p^\mu },$ the number density of bosons with four-momentum $p^\mu ,$ we require that
\beq
m-\mu -\mu _K\frac{m^2}{2M}\geq 0.
\eeq
The discriminant for the l.h.s. of the inequality must be nonnegative,
i.e.,
\beq
\mu \leq \frac{M}{2\mu _K}.
\eeq
For such $\mu ,$ (2.2) has the solution
\beq
m_1\equiv \frac{M}{\mu _K}\left( 1-\sqrt{1-\frac{2\mu \mu _K}{M}}\right) \leq
m\leq
\frac{M}{\mu _K}\left( 1+\sqrt{1-\frac{2\mu \mu _K}{M}}\right) \equiv m_2.
\eeq
For small $\mu \mu _K/M,$ the region (2.4) may be approximated by
\beq
\mu \leq m\leq \frac{2M}{\mu _K}.
\eeq
One sees that $\mu _K$ plays a fundamental role in determining an upper
bound of the mass spectrum, in addition to the usual lower bound $m\geq
\mu .$ In fact, small $\mu _K$ admits a very large range of off-shell
mass, and hence can be associated with the presence of strong
interactions \cite{MS}. For our present purposes it will be sufficient
to assume that the mass distribution has a finite range  $m_1\leq
m\leq m_2$ around the on-shell value $m_c= M/\mu_K$ corresponding to the 
limiting value for which the inequality (2.3) becomes an equality.

In order to show that our results hold independent of the dimensionality of spacetime, we shall consider our ensemble in 
one temporal and $D$ spatial dimensions, $D\geq 1.$

Replacing the sum over $p^\mu $ (2.1) by an integral, $$\sum _{p^\mu }\Longrightarrow \frac{V^{(1+D)}}{(2\pi )^{1+D}}\int d^{1+D}p,$$
where $V^{(1+D)}$ is the system's $(1+D)$-volume, and using the relation $(p^\mu =(p^0,{\bf p}))$ $$d^{(1+D)}p=\frac{d^Dp}{2p^0}dm^2,\;\;m^2\equiv -p^\mu p_\mu ,\;\;\mu =0,1,\ldots ,D,$$ one obtains for the density of events per unit $(1+D)$-volume, $n\equiv N/V^{(1+D)},$
\beq
n=\int _{m_1}^{m_2}\frac{dm\;m}{2\pi }\int \frac{d^D{\bf p}}{(2\pi )^Dp^0}f(p),
\eeq
with $f(p)\equiv n_{p^\mu },$ as given in Eq. (2.1). 
Typical average values are given by the relations
\beq
\langle p^\mu \rangle =\frac{1}{n}\int _{m_1}^{m_2}\frac{dm\;m}{2\pi }
\int \frac{d^D{\bf p}}{(2\pi )^Dp^0}\;p^\mu f(p),
\eeq
\beq
\langle p^\mu p^\nu  \rangle =\frac{1}{n}\int _{m_1}^{m_2}\frac{dm\;m}{2\pi }
\int \frac{d^D{\bf p}}{(2\pi )^Dp^0}\;p^\mu p^\nu f(p),\;\;\;{\rm etc.}
\eeq

To find the expressions for the pressure and energy density in our ensemble, we study the particle energy-momentum tensor defined by the relation\footnote{The corresponding relation of ref. \cite{HSS} is given in four-dimensional spacetime.} \cite{HSS}
\beq
T^{\mu \nu }(q)=\sum _i\int d\tau \frac{p_i^\mu p_i^\nu }{m_c}
\delta ^{1+D}(q-q_i(\tau )),
\eeq
in which $m_c$ is the value around which the mass of the events making up the 
ensemble is distributed. Upon integrating over a small $(1+D)$-volume 
$\triangle V$ and taking the ensemble average, (2.9) reduces to \cite{HSS}
\beq
\langle T^{\mu \nu }\rangle =\frac{T_{\triangle V}}{m_c}n\langle
p^\mu p^\nu \rangle .
\eeq
In this formula, $n=N/V$, and $T_{\triangle V}$ is the average passage interval in
$\tau $ for the events which pass through $\triangle V,$ which we discussed above. The formula (2.10) reduces, through Eq. (2.8), to
\beq
\langle T^{\mu \nu }\rangle =\frac{T_{\triangle V}}{2\pi m_c}\int _{m_1}^{m_2}dm\;m\int \frac{d^D{\bf p}}{(2\pi )^Dp^0}\;p^\mu p^\nu f(p).
\eeq
Using the standard expression
\beq
\langle T^{\mu \nu }\rangle =pg^{\mu \nu }-(p+\rho )u^\mu u^\nu ,
\eeq
where $p$ and $\rho $ are the particle pressure and energy density,
respectively, we obtain $$\rho =\langle T^{00}\rangle,\;\;\;p=\frac{1}{D}g^{ii}\langle T_{ii}\rangle ,\;\;i=1,2,\ldots, D,$$ and therefore, through (2.11),
\beq
p=\frac{T_{\triangle V}}{2\pi m_c}\int _{m_1}^{m_2}dm\;m\int \frac{d^D{\bf p}}{(2\pi )^D}\frac{{\bf p}^2}{Dp^0}f(p),
\eeq
\beq
\rho =\frac{T_{\triangle V}}{2\pi m_c}\int _{m_1}^{m_2}dm\;m\int \frac{d^D{\bf p}}{(2\pi )^D}\;p^0 f(p).
\eeq
We now calculate the particle number density per
unit $D$-volume. The particle $D+1$-current is given by the
formula \cite{HSS}
\beq
J^\mu (q)=\sum _i\int d\tau \frac{p^\mu _i}{m_c}\delta ^{1+D}(q-q_i(\tau )),
\eeq
which upon integrating over a small $(1+D)$-volume and taking the
average reduces to
\beq
\langle J^\mu \rangle =\frac{T_{\triangle V}}{m_c}n\langle p^\mu \rangle ;
\eeq
then the {\it particle} number density \cite{Synge,Hakim} is
\beq
N_0\equiv \langle J^0\rangle =\frac{T_{\triangle V}}{m_c}n\langle E\rangle ,
\eeq
so that, with the help of Eq. (2.7),
\beq
N_0=\frac{T_{\triangle V}}{2\pi m_c}\int _{m_1}^{m_2}dm\;m\int \frac{d^D{\bf p}}{(2\pi )^D}\;f(p).
\eeq
Since
\beq 
p=\int \frac{d^D{\bf p}}{(2\pi )^D}\frac{{\bf p}^2}{Dp^0}f(p)\equiv p(m),
\eeq
\beq
\rho =\int \frac{d^D{\bf p}}{(2\pi )^D}\;p^0 f(p)\equiv \rho (m)
\eeq
and
\beq
N_0=\int \frac{d^D{\bf p}}{(2\pi )^D}\;f(p)\equiv N_0(m)
\eeq
are the standard expressions for the pressure, energy density and particle number density in $1+D$ dimensions, respectively 
\cite{HW,Act,1+D}, we have the following relations:
\beq
p=\frac{T_{\triangle V}}{2\pi m_c}\int _{m_1}^{m_2}dm\;m\;p(m),
\eeq
\beq
\rho =\frac{T_{\triangle V}}{2\pi m_c}\int _{m_1}^{m_2}dm\;m\;\rho (m),
\eeq
\beq
N_0=\frac{T_{\triangle V}}{2\pi m_c}\int _{m_1}^{m_2}dm\;m\;N_0(m).
\eeq
It is seen in these relations that the manifestly covariant framework
provides {\it a linear mass spectrum, independent of the dimensionality of spacetime}. In order to obtain the expressions for the basic thermodynamic quantities, one has to integrate the standard (on-shell) results over this spectrum within the range of the mass distribution. 

Using the formulas (2.22)-(2.24), one can establish the uncertainty relation (1.10) for a narrow mass width around the on-shell value: as $m\rightarrow m_c,$ 
\beq
\int _{m_1}^{m_2}dm\;m\;f(m)\rightarrow \triangle m\;m_c\;f(m_c),
\eeq
where $f(m)$ stands for each of the $p(m),\rho (m),N_0(m),$ and $\triangle m$ is the (infinitesimal) width of the mass distribution around $m_c.$ The requirement that the results for $p,\rho $ and $N_0$ coincide with those of the usual on-shell theories implies $p=p(m_c),\rho =\rho (m_c),N_0=N_0(m_c)$ in Eqs. (2.22)-(2.24), which leads, with Eq. (2.25), to the relation\footnote{In c.g.s. units, this relation has a factor $\hbar /c^2$ on the right hand side.}
\beq
T_{\triangle V}\triangle m=2\pi ,
\eeq
the case of equality in the relation (1.10),
in agreement with the results obtained earlier in ref. \cite{glim}.
One can understand this relation, up to a numerical factor, in terms of
the uncertainty principle (rigorous in the $L^2(R^4)$ quantum theory)
$\triangle E\cdot \triangle t \geq 1/2.$ Since the time
interval for the particle to pass the volume $\triangle V$ (this smallest
macroscopic volume is bounded from below by the size of the wave packets)
$\triangle t\simeq E/M \;\triangle \tau ,$ and the dispersion of $E$ due
to the mass distribution is $\triangle E\sim m\triangle m/E,$ one obtains
a lower bound for $T_{\triangle V}\triangle m$ of order unity.

\section{Mass-proper time uncertainty relation}
We now wish to prove the relation (1.10) for the general case, not
only for the case of a narrow mass width as done in the previous
section. First, we note that we previously considered a relativistic
ensemble without degeneracy; therefore no degeneracy factor appeared
in the expressions for the basic thermodynamic quantities. Now suppose
that we have $\nu$ internal degrees of freedom in our ensemble which correspond
to degeneracy. In this case considerations made previously will remain valid and lead to the formulas (2.22)-(2.24) with the extra factor of $\nu$ on their right hand side:
\beq
p=\frac{T_{\triangle V}\nu}{2\pi m_c}\int _{m_1}^{m_2}dm\;m\;p(m),
\eeq
\beq
\rho =\frac{T_{\triangle V}\nu}{2\pi m_c}\int _{m_1}^{m_2}dm\;m\;\rho (m),
\eeq
\beq
N_0=\frac{T_{\triangle V}\nu}{2\pi m_c}\int _{m_1}^{m_2}dm\;m\;N_0(m).
\eeq
On the other hand, according to our previous arguments,
one can consider the $\nu$ degrees of freedom as being distributed in the mass interval $m_1\leq m\leq m_2$ with a linear mass spectrum, 
\beq 
\tau (m)=Cm,
\eeq
which leads to the formula
\beq
p=\int _{m_1}^{m_2}dm\;\tau (m)\;p(m),
\eeq
and analogous relations for $\rho $ and $N_0$ (similar to the
treatment of a strongly interacting system by means of a particle
resonance spectrum \cite{Shu}). In fact, the linear mass spectrum finds its confirmation in the experimental hadronic resonance spectrum: if one calculates, for example, the pressure in the hadronic resonance gas by summing up the individual contributions of a finite number of the different hadronic species with the corresponding degeneracies, $$p=\sum _ig_i\;p(m_i),\;\;\;p(m_i)=\frac{T^2m_i^2}{2\pi ^2}K_2(m_i/T),$$ and by using the formula (3.5), in which $m_1$ and $m_2$ are the masses of the lightest and the heaviest species, respectively, the results coincide \cite{linear}.

The normalization constant $C$ is determined by the condition
\beq
\int _{m_1}^{m_2}\tau (m)\;dm=\nu;
\eeq
therefore 
\beq
C=\frac{2\nu}{m_2^2-m_1^2}=\frac{\nu}{\triangle m\;(m_1+m_2)/2},
\eeq
where $\triangle m=m_2-m_1$ is the width of the mass distribution in a general case. Since $m_c$ should be associated with one of the averages $\langle m\rangle $ or $\langle m^2\rangle $ which both are closer to $m_2$ than to $m_1$ for a linear spectrum [$(m_1+m_2)/2\stackrel{<}{\sim } m_c$], 
\beq
C\stackrel{>}{\sim }\frac{\nu}{m_c\;\triangle m}.
\eeq

Direct comparison of the formulas (3.1)-(3.3) with the relation (3.5) and analogous formulas for $\rho $ and $N_0,$ with $\tau (m)$ given by (3.4), 
\beq
p=C\int _{m_1}^{m_2}dm\;m\;p(m),
\eeq
\beq
\rho =C\int _{m_1}^{m_2}dm\;m\;\rho (m),
\eeq
\beq
N_0=C\int _{m_1}^{m_2}dm\;m\;N_0(m),
\eeq
leads to the relation $$C=\frac{T_{\triangle V}\nu}{2\pi m_c},$$ which reduces,
through (3.8), to
\beq
T_{\triangle V}\triangle m\stackrel{>}{\sim }2\pi ,
\eeq 
the relation (1.10) for the general case of the finite range mass distribution.

In order that our considerations be valid, the effective degeneracy in
a relativistic ensemble should be large, $\nu >>1,$ so that one could
speak of the distribution of $\nu$ degrees of freedom in a finite mass
range. This is the case for realistic physical systems such as high temperature strongly interacting hadronic matter \cite{Shu}. 

\section{Concluding remarks}
In this paper we have proved the uncertainty relation (1.10) for the
general case of a finite range of mass distribution in a relativistic
ensemble. This relation allows one to admit the following picture of a
strongly interacting system: one can consider a strongly interacting
system as a distribution of free particles which  temporarily go off-shell
while undergoing an interaction. Then $T_{\triangle V}$ may be
associated with the time for the particle to remain very close to its mass shell. So, for weakly interacting systems, $\triangle m\simeq 0,$ $T_{\triangle
V}\simeq \infty ,$ i.e., the particle remains on-shell almost all the
time. In contrast, for a strongly interacting system, $\mu _K\simeq 0$
\cite{MS}; then, in view of (2.5), $\triangle m\simeq \infty ,$ and
$T_{\triangle V}\simeq 0,$ i.e., the particle is off-shell almost
always (because it undergoes interaction almost continuously).

We have remarked that the non-relativistic limit of the mass-``proper
time'' uncertainty relation provides a new derivation of the $\triangle
E \triangle t  \stackrel{<}{\sim } \hbar $ relation of the
non-relativistic theory.  This result implies, as we pointed out, the
existence of a residual ensemble over $t$, even in the
non-relativistic limit. Such an ensemble has been introduced recently
\cite{lax} in order to achieve an exact semigroup (exponential decay)
law of evolution for the reduced motion of an unstable system.  The
usual derivation\cite{jammer} of the non-relativistic energy-time uncertainty
relation studies the motion of the system under the action of the full
Hamiltonian relative to the eigenstates of unperturbed energy; this
procedure corresponds precisely to that of Weisskopf and
Wigner\cite{wigner} for the description of the decay of unstable
systems. It is argued in refs. \cite{lax} that the introduction of an
ensemble over $t$ is necessary for achieving the semigroup property as
well as for the consistency of the interpretation.

The uncertainty relation (1.10) may be useful in practical calculations concerning realistic strongly interacting systems in which the particles are necessarily off-shell. For example, it may allow one to estimate the relaxation times for the quark-gluon plasma created in ultrarelativistic heavy-ion collisions.


\bigskip
\bigskip
\begin{thebibliography}{9}
\bibitem{Sudarshan} D.G. Currie, J. Math. Phys. {\bf 4} (1963) 470; D.G.
Currie, T.F. Jordan and E.C.G. Sudarshan, Rev. Mod. Phys. {\bf 35} (1963) 350
\bibitem{NW} T.D. Newton and E.P. Wigner, Rev. Mod. Phys. {\bf 21} (1949) 400
\bibitem{AH} R. Arshansky and L.P. Horwitz, Found. Phys. {\bf 15} (1985) 701
\bibitem{Shnerb} N. Shnerb and L.P. Horwitz, Phys. Rev. A {\bf 48} (1993) 4068
\bibitem{Synge} J.L. Synge, {\it The Relativistic Gas,} (North-Holland,
Amsterdam, 1957)
\bibitem{Pauli} W. Pauli, {\it Theory of Relativity,} (Pergamon, Oxford, 1958)
\bibitem{HSP} L.P. Horwitz, W.C. Schieve and C. Piron,
Ann. Phys. (N.Y.)  {\bf 137} (1981) 306
\bibitem{BH1} L. Burakovsky and L.P. Horwitz, Physica A, {\bf 201} (1993) 666
\bibitem{Kap1} J.I. Kapusta, {\it Finite-Temperature Field Theory,} (Cambridge
University Press, Cambridge, 1989)
\bibitem{Stu} E.C.G. Stueckelberg, Helv. Phys. Acta {\bf 14} (1941)
372, 588; {\bf 15} (1942) 23
\bibitem{HP} L.P. Horwitz and C. Piron, Helv. Phys. Acta {\bf 46} (1973) 316
\bibitem{AHL} R. Arshansky, L.P. Horwitz and Y. Lavie, Found. Phys.
{\bf 13} (1983) 1167
\bibitem{HSS} L.P. Horwitz, S. Shashoua and W.C. Schieve,
Physica A  {\bf 161} (1989) 300
\bibitem{FW} J.R. Fanchi and W.J. Wilson, Found. Phys. {\bf 13} (1983) 571
\bibitem{LP} L. Landau and R. Peierls, Z. Phys. {\bf 69} (1931) 56
\bibitem{MS} D.E. Miller and E. Suhonen, Phys. Rev. D {\bf 26} (1982) 2944
\bibitem{Hakim} R. Hakim, J. Math. Phys. {\bf 8} (1967) 1315.
%\bibitem{degen} L. Burakovsky, L.P. Horwitz and W.C. Schieve, Statistical Mech%anics of Relativistic Degenerate Fermi Gas I.Cold Adiabatic Equation of State,% Tel-Aviv Univ. Preprint TAUP-2124-93; unpublished
%\bibitem{realeq} L. Burakovsky, L.P. Horwitz and W.C. Schieve, On the Relativi%stic Statistical Mechanics of Strongly Interacting Matter, Tel-Aviv Univ. Prep%rint TAUP-2231-95; unpublished \\ See also L. Burakovsky, L.P. Horwitz and W.C%. Schieve, A New Relativistic High Temperature Bose-Einstein Condensation; to %appear in Phys. Rev. D
\bibitem{HW} H.A. Haber and H.E. Weldon, J. Math. Phys. {\bf 23} (1982) 1852
\bibitem{Act} A. Actor, Nucl. Phys. B {\bf 256} (1986) 689
\bibitem{1+D} L. Burakovsky, On Relativistic Statistical Mechanics, Thermodynamics and Cosmology in 1+D Dimensions, Tel-Aviv Univ. Preprint TAUP-2187-94;
unpublished
\bibitem{glim} L. Burakovsky and L.P. Horwitz, J. Phys. A {\bf 27} (1994) 4725
\bibitem{Shu} E.V. Shuryak, {\it The QCD Vacuum, Hadrons and the Superdense Matter,} (World Scientific, Singapore, 1988)
\bibitem{linear}  L. Burakovsky, L.P. Horwitz and W.C. Schieve,
Hadronic Resonance Spectrum: Power vs. Exponential Law, Experimental
Evidence. In preparation.
\bibitem{lax} L.P. Horwitz and C. Piron, Helv. Phys. Acta {\bf 66},
693(1993); E. Eisenberg and L.P. Horwitz, {\it Time,
irreversibility and the unstable system in quantum and classical
physics}, to be published in Advances in Chemical Physics (1996); see
also L.P Horwitz and E. Eisenberg, in {\it Proceedings of the Conference on
Quantum-Classical Correspondence}, Drexel University, 8 September,
1994.
\bibitem{jammer} M. Jammer, {\it The Conceptual Development of Quantum
Mechanics}, McGraw Hill, New York (1966).
\bibitem{wigner} V.F. Weisskopf and E.P. Wigner, Zeits, f, Phys. {\bf
63}, 54 (1930); {\bf 65}, 18 (1930).
\end{thebibliography}
\end{document}